# The nature of the diffuse light near cities detected in nighttime satellite imagery

**Alejandro Sánchez de Miguel**[1,2,3*], **Christopher C. M. Kyba**[4,5], **Jaime Zamorano**[2], **Jesús Gallego**[2], **and Kevin J. Gaston**[1]

[1]Environment and Sustainability Institute, University of Exeter, Penryn, Cornwall TR10 9FE, UK
[2]Dept. Física de la Tierra y Astrofísica and Instituto de Física de partículas y del Cosmos IPARCOS, Universidad Complutense de Madrid, Madrid, 28040, Spain
[3]Instituto de Astrofísica de Andalucía, Glorieta de la Astronomía, s/n,C.P.18008 Granada, Spain
[4]GFZ German Research Centre for Geosciences, 14473 Potsdam, Germany
[5]Leibniz-Institute of Freshwater Ecology and Inland Fisheries, 12587 Berlin, Germany
[*]a.sanchez-de-miguel@exeter.ac.uk

## ABSTRACT

Diffuse glow has been observed around brightly lit cities in nighttime satellite imagery since at least the first publication of large scale maps in the late 1990s. In the literature, this has often been assumed to be an error related to the sensor, and referred to as "blooming", presumably in relation to the effect that can occur when using a CCD to photograph a bright source. Here we show that the effect is not instrumental, but in fact represents a real detection of light scattered by the atmosphere. Data from the Universidad Complutense Madrid sky brightness survey are compared to nighttime imagery from multiple sensors with differing spatial resolutions, and found to be strongly correlated. These results suggest that it should be possible for a future space-based imaging radiometer to monitor changes in the diffuse artificial skyglow of cities.

## Introduction

A diffuse glow of light surrounding cities, such as that seen in Figure 1, was noted by the Earth Observation Group (EOG) of the National Oceanic and Atmospheric Administration (NOAA) from their earliest publications of nighttime light data sets (e.g.[1,2]). Due to the low spatial resolution of the Defence Meteorological Satellite Program Operational Linescan System (DMSP), it was not possible to determine the nature of these observations, and in particular whether they were instrumental errors or real observations of light. Since that time, many publications have referred to this glow as "blooming", perhaps by analogy to the effect seen near bright sources in early CCD imagery (e.g.[1,3]). This glow presents a problem for some analyses (e.g. detection of urban areas) regardless of its cause, and has therefore frequently been effectively treated as an instrumental error. Newer satellite data are no longer restricted by the saturation problems and low spatial resolution of DMSP data, but still contain the glow near cities (Figure 1). This strongly suggests that the glow is not an instrumental effect, but rather real light being detected by the instruments.

Observers from the ground have long been aware of another type of glow near cities, the artificial sky brightness known as "skyglow", a form of light pollution.[5–7] Skyglow is caused when light that is emitted upward (or reflected from the ground) scatters off molecules or aerosols in Earth's atmosphere, and is re-directed towards the ground.[8] Systematic academic study of skyglow began in the 1970s,[9–12] and physical modeling of skyglow began in the 1980s;[13, 14] we recently learned that the first experimental studies of skyglow were conducted by the US War Department during World War 2.[15] Could the diffuse glow around cities in satellite imagery be caused by the same effect?

Figure 2 demonstrates how scattered light is attributed to an incorrect location in observations of night lights from space. The scattered light appears to come from a position on the line of sight, rather than from its true emission location. This effect could therefore explain, for example, the detection of light seeming to emanate from Lake Michigan in regions near Chicago (Figure 1). But how can we be sure that this is the cause, and if it is, what is the relationship between the skyglow observed from the ground and that observed from space?

Rayleigh scattering has both forward-backward and azimuthal symmetry, so the fraction of light scattered back towards Earth by molecules must be nearly equal to the amount scattered upwards (ignoring a minor difference caused by the curvature of the Earth). Aerosols primarily scatter light through small angles in the forward direction. For upward directed light in general, one would therefore expect a difference in their contribution to skyglow observed from space compared to the ground. However, the main contributor to skyglow observed from the ground is light that is emitted at near-horizontal angles.[8,15,16] For horizontally directed light, the azimuthal symmetry of Mie scattering implies that a similar amount of large angle scatter would occur in both the upward and downward direction. Taken together, these facts imply that if atmospheric scattering is the cause of the glow around cities, one must expect a fairly strong relationship between skyglow observed from the ground and the glow in satellite imagery.

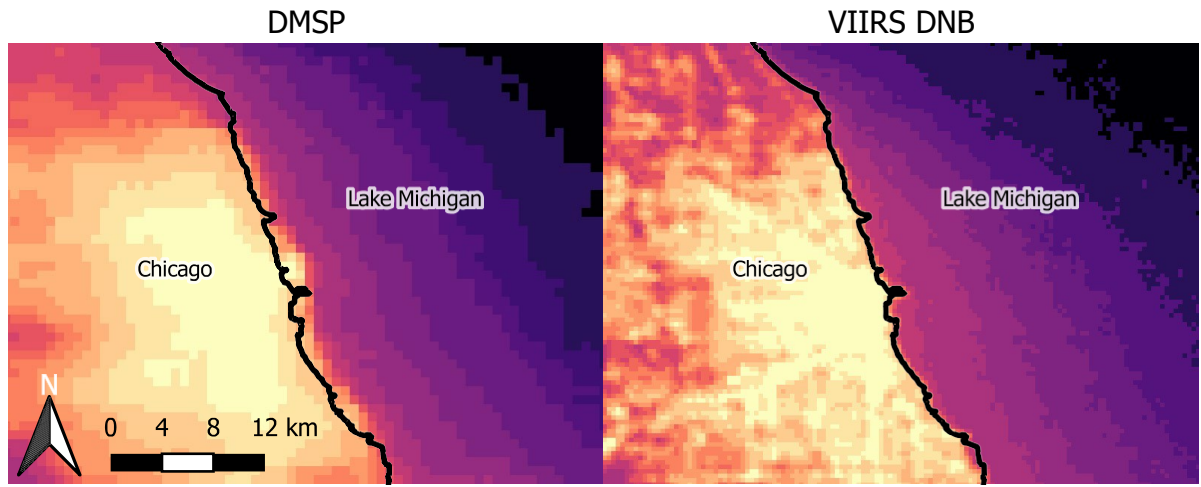

**Figure 1.** Images of the area near the city of Chicago, Illinois, using EOG's radiance calibrated DMSP composite from 2010 (left) and the DNB composite from October 2012 (right). The black line marks the shore of the lake. Note that the glow extends well out onto Lake Michigan (right side), an area with little or no light emission. The color scales are both logarithmic, but are not identical (DMSP and DNB data cannot be directly compared). See also Figures 10 and 12 in Levin 2017.[4]

Kyba et al. and Zamorano et al. have previously examined the relationship between ground based observations of skyglow and the DMSP data.[17, 18] Much of the data used in Kyba et al. were from lit locations, and they dismissed the possibility that the DMSP values directly represented skyglow. Zamorano et al., on the other hand, took data primarily from locations with no installed lighting, and explicitly raised the possibility of this connection.[17] In this work, we use data taken in areas away from installed lighting to test directly the relationship between skyglow observed from the ground and from space. In addition to those from the DMSP, data with higher resolution and sensitivity from the DNB and astronaut photographs are also analyzed.[19–22]

## Results

There is a strong correlation between the zenith night sky radiance data from the UCM Sky brightness survey (see Methods) and the space-based datasets with both low (Fig. 3) and high resolution (Fig. 4). Sky radiance is shown in $mag_{SQM}/arcsec^2$, an astronomical unit for which larger values represent darker skies (see Methods, Puschnig et al.,[23, 24] and Sanchez de Miguel et al.[25]). Note that the radiance observed by DMSP and photographs from the International Space Station (ISS) cannot be directly compared, because of the different spectral sensitivity of the instruments.[26] While the bulk of the data are well correlated in both cases, in the high resolution data some sky brightness observations were considerably darker than expected, given the radiance in the ISS imagery. These likely represent cases where pixels included direct light emissions in addition to the scattered light from the city of Madrid and its surroundings.

For each space-based observation, we performed a least squares regression to predict the (logarithmic) Sky Quality Meter derived radiance ($mag_{SQM}/arcsec^2$) based on the logarithm of the radiance observed by satellite (in $nW/cm^2sr$). With this fit, it is then possible to compare the individual observations to the prediction to examine the dispersion of the data. The fit residuals (predicted minus observed radiance in $mag_{SQM}/arcsec^2$) are shown for each of the four space-based observations in Figure 5. The satellite observations have an intrinsic resolution of 5 km (DMSP), 750 m (DNB), 240 m (ISS040-E-08258-62), and 54 m (ISS030-E-188208-10).

It is clear that for the high resolution imagery data, there is a set of data on the high side, meaning that the sky is darker than we would expect based on the space-based image. This data is fairly well described by a double Gaussian fit (e.g. Figure 6). The half width at half maximum of the main component (taller left hand peak) of the double Gaussian fit to the residual histograms is about $\sim 0.35$ $mag_{SQM}/arcsec^2$ for each of the space-based datasets. In terms of radiance, this corresponds to a typical difference of roughly ±35%. We believe that the second Gaussian distribution is caused by pixels which contain a mixture of both upward scattered skyglow and direct light emissions from the ground, whereas the first Gaussian represents

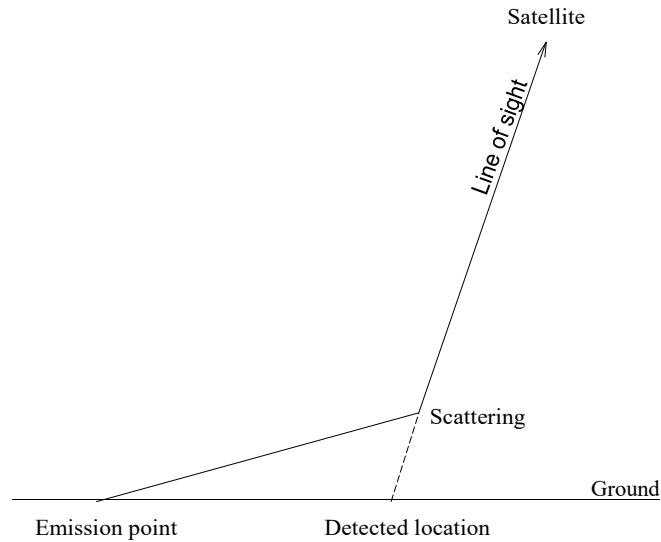

**Figure 2.** Schematic diagram showing how scattered light is falsely attributed to a location in nighttime satellite imagery. Light emitted from the emission point undergoes a scattering, changing its direction. The satellite radiometer viewing along the direction that the light was propagating attributes it to a false location.

pixels which contain only upward scattered skyglow.

These results suggest that it should be possible to create regional sky brightness (see fig.11) or even global sky brightness maps based on real VIIRS measurements, as is indeed needed according to theoretical considerations.[27] Using Google Earth Engine[28] a world map has been created that can be examined at Google Earth Engine, at URL: https://pmisson.users.earthengine.app/view/trends. The current version will be improved in the future, once new and better calibrations of VIIRS and sky brightness databases are made publicly accessible.

### Relationship between sky brightness and diffuse light

Data from the UCM Sky brightness survey were compared with data from the ISS (two ISS HDR composites), an image from the VIIRS (May 2014) and an image from the DMSP (2011); data from the VIIRS and the DMSP were provided by the EOG/NOAA group. We found correlations between all of these data sets, but the spatial resolutions are very different. The mosaic of ISS030-E-188206-10 had a resolution (PSF) of 54 m, that of ISS040-E-08258-62 had a resolution of 240 m, the VIIRS data had a resolution of 750 m, and the DMSP data a resolution of 5 km. The UCM Sky brightness survey took linear measurements with a precision of better than 10 m. All of these data were re-sampled on standard grids of 15, 30 and 60 arcsec, with the exception of those from the DMSP, because of its own large PSF. It was already known that there was a relationship between sky brightness and images from the DMSP.[18,29]

The correlation in regions with no direct illumination, with similar dispersion, between four different image systems leads us to conclude that the diffuse light observed in these regions has the same source, sky brightness, dispersed upwards. The changes in fit are produced by the different mean composition of the atmosphere when the image was taken. This fact can be used to produce sky brightness maps based on satellite data, for example as in Fig. 11 for the Madrid region. Another factor arguing for the real existence of this diffuse emission is the relationship between the diffuse emission to the amount of aerosols.[30]

### Comparison with the World Sky atlas

Another way to verify the accuracy of this VIIRS data is to compare it with the predictions of a Sky Brightness model[31] based on the Madrid Sky brightness survey and other measurements. To do so, we selected a region with no interference of direct lights and low albedo, namely the sea surrounding the region of Valencia and the Balearic Islands. To increase the signal to noise of the VIIRS data we used the median of the months closest to the four months second equinox (Aug to Nov), and then average the values from 2012 to 2018. As a result, we find a perfect fit between the model predictions and the observed values (Figure 13.

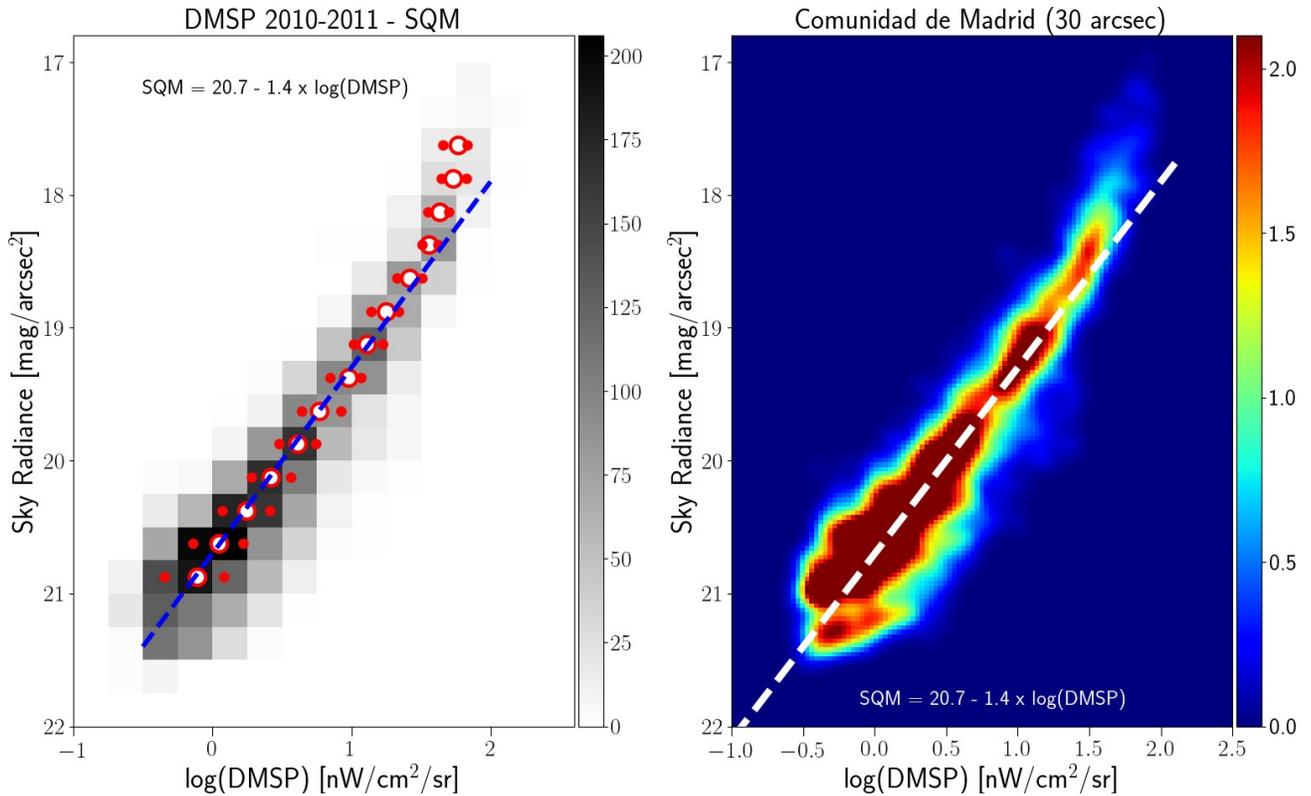

**Figure 3.** Correlation between the sky brightness at zenith (UCM sky brightness survey) and the DMSP 2011 radiance calibrated data of the region of Madrid. The SQM data have been averaged on the same 30 arcsec grid as the DMSP data. The left hand plot is a 2D histogram of the radiances as measured from Earth and from space. The number of 30 arcsec cells for each bin is coded as gray shades. The open dots and smaller red dots represent the median value for each sky radiance bin and the 25% and 75% percentile respectively. The dashed blue line is a linear fit using the darkest values and rejecting the bins brighter than 19 mag/arcsec$^2$. The right hand panel shows the same result using a non-binned color coded density plot with a Gaussian filter. The dashed white line reproduces the fit of the left panel and has been added for comparison purposes.

## Discussion

The diffuse light around cities on nighttime satellite imagery of the earth has often been interpreted as instrumental error. Here we have shown that this is not the case. Rather, this light is mostly due to sky brightness propagated from urban areas.

The variation between observed and DMSP predicted sky brightness can be compared to that previously published by Kyba et al.[18] In Kyba et al., the standard deviation of the residuals was 0.5 mag$_{SQM}$/arcsec$^2$, whereas here we observed 0.3 mag$_{SQM}$/arcsec$^2$. A major difference between the two works is that Kyba et al. was based on observations submitted in a citizen science project. The improved accuracy in this paper can be attributed to several factors. First, the UCM sky brightness survey was performed by a single group of researchers using a single protocol and in similar weather conditions. In contrast, the citizen science data may include data from more turbid atmospheres, and the overly bright outliers in the citizen science data suggest that it was often taken too close to light sources. Also, the data from the UCM survey have been very carefully filtered to avoid false values. This issue is more difficult to address in a citizen science project.

It is possible to see from fig 5 that the dispersion of the main component is similar for all the satellites. The main difference between them is the shape of the tail to greater brightness. In the case of the DMSP, this tail it is not very pronounced because of the blurriness of its Point Spread Function(PSF). In the case of the other satellites, it's clear that the distribution is more asymmetric. The interpretation of this behavior is clear when we change the size of the sampling. In Figure 5, it is possible to see how when we increase the size of a pixel, the tail begins to grow, and the size of the main peak decreases. This effect is produced due to the admixture of direct illuminated and diffuse illuminated signal within the same pixel. The sky brightness is more strongly related to the diffuse light than with the direct illumination, so when resolution is reduced, information is lost. In Figure 4 we can see how it is possible to reproduce the effect of loss of resolution by increasing the pixel size.

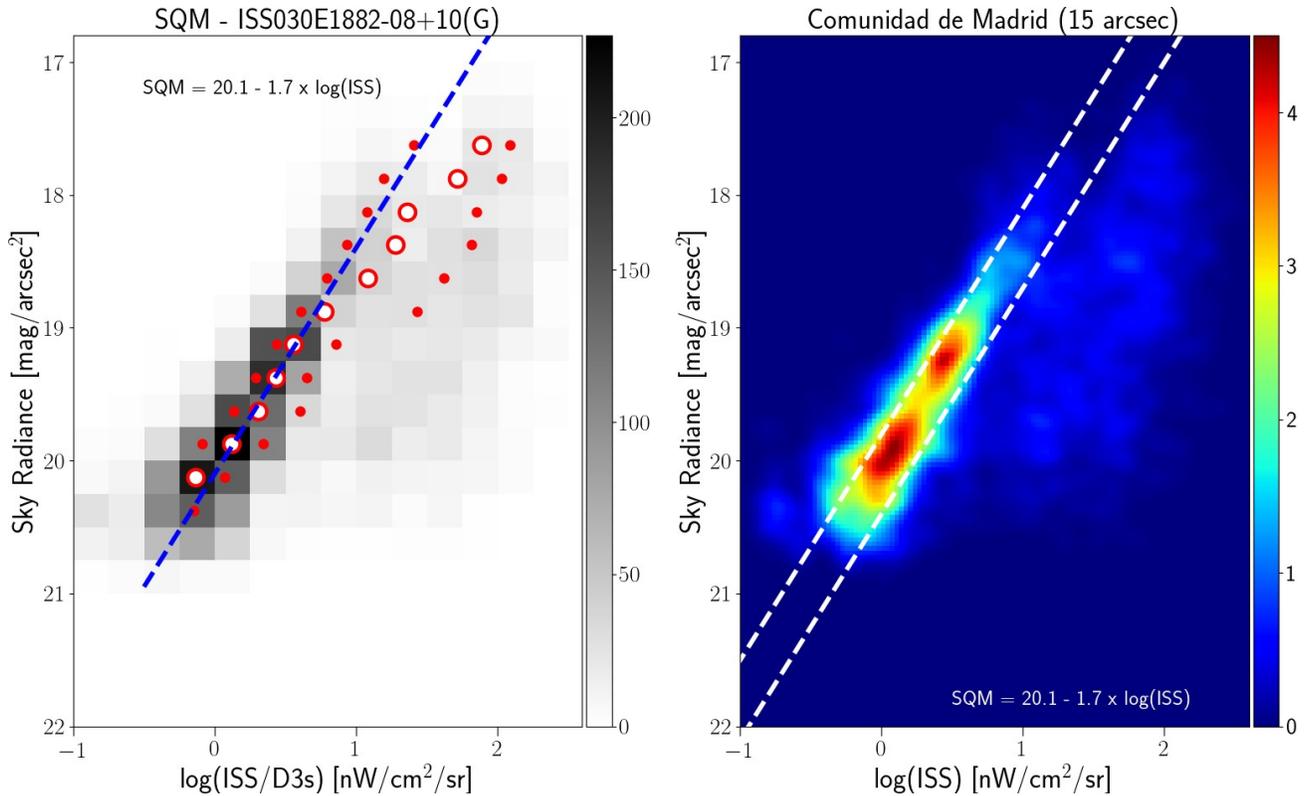

**Figure 4.** Correlation between the sky brightness at zenith (UCM sky brightness survey) and a composite high dynamic range image produced from astronaut photographs (ISS030E188208 and ISS030E188210) of the region of Madrid. The SQM individual data and the ISS image have been averaged on the same 15 arcsec grid. The left hand plot is a 2D histogram of the radiances as measured from earth and from space. The number of 15 arcsec cells for each bin is coded as gray shades. The open dots and smaller red dots represent the median value for each sky radiance bin and the 25% and 75% percentile respectively. The dashed blue line is a linear fit using the darkest values and rejecting the bins brighter than 18.8 mag/arcsec$^2$. The right hand panel shows the same result using a non-binned color coded density plot with a Gaussian filter. We have added two dashed white lines with the same slope as the blue line in the left panel for comparison purposes.

We can conclude that the radiance observed in the regime between 0.2-5 $nW/cm^2/sr$ is dominated by diffuse light comes from light scattered in the atmosphere to the surroundings of Madrid. These values need to be inspected in each location, as dim direct light can be also occur in that regimen for small towns. As the Madrid data are representative of sky brightness in the rest of the world, this approach can be extrapolated more widely, although values will need to be adjusted locally because the typical brightness and spectra of cities can change dramatically. A way of removing the sky glow effect is to subtract the sky brightness of Falchi et. al. 2016.[31] Other components such as natural sky glow (auroras and airglow), reflectance of the ground, fogs and transients need to be studied in more depth in other to extract all the information of the VIIRS imagery and new sky brightness surveys are needed in areas sensitive to effects not considered by Falchi et. al. 2016, such as blocking effects of orthography.

### Other components of the diffuse light: albedo, natural airglow, sea fogs and real blooming

It is beyond the scope of this article to address all of the components of the dim diffused light present in VIIRS images. However, not all of the diffuse light can be explained by sky brightness effects. Albedo or reflectance, natural airglow and aurora, sea fogs and some real blooming can be detected in VIIRS images. Indeed, the sensitivity of the VIIRS camera is so high that it can even detect the reflectance of starry skies on the ground. Fig. 14 shows the correlation between the median of the Albedo bands of MODIS and the VIIRS for the months closest to the four second equinox (Aug to Nov), and averaged from 2012 to 2018, for the region surrounding the Wal al Namus volcano, one of the most extreme changes of albedo on the surface of the Earth. In this case, the correlation detected is also clear ($R^2 = 0.93$), although there is some structure that indicates that the spectral sensitivities of the VIIRS and MODIS do not match perfectly, so this result might be improved. This effect was previously reported by Roman et. al 2018.[32]

The atmosphere naturally emits line emissions that can be detected from the ground. These also contribute to sky brightness but in a highly variable fashion. Fog at sea can also contribute to diffuse light in nighttime satellite imagery. Because it can have the same temperature as the sea itself, this effect can be challenging to remove. The reflectance of the sea is, however, lower than that of fog, giving a brighter signal in the middle of the sea, mainly in the North Atlantic and North Pacific oceans and around the Equator.

Finally, coming full circle, although the majority of diffuse light in nighttime imagery arises from sky brightening and not blooming, a component of blooming does remain. This is detectable in the high gain regime of VIIRS as tails to the distribution of brightness values that are a consequence of long exposure or blooming from the window of the instrument. This effect can be clearly seen in figure 15.

## Methods

### Physical base

All current and previous models of sky brightness are based on Rayleigh and Mie scattering.[13, 33–37] The main contributor to the brightness of the sky is direct light emissions at a low angle.[8,15,16] Because of the up/down symmetry of Rayleigh and Mie diffusion for light traveling horizontally, it is reasonable to suppose that the same amount of light that is scattered to the ground from the zenith is also scattered towards space from the same volume of atmosphere.

From 2009 to 2014, the UCM group undertook a sky brightness survey around the Madrid region with Sky Quality Meter (SQM[38–40]) photometers mounted onboard cars. Taken under different atmospheric conditions, this has produced more than 30,000 individual measurements. When the Region of Madrid is divided into cells of 2.2 km$^2$, 60% of the total area is covered.[17,22,41] During this period there was no significant change in the overall street lighting power consumption in Madrid.[42]

In order to test the relationship between the diffuse light seen from the satellite and skyglow detected from the ground, we used all of the data available. The data from the SNPP/VIIRS/DNB and the DMSP/OLS are publicly available and produced by the Earth Observation Group. The Skyglow and DMSP of the Madrid area were discussed in depth in Zamorano et al.[43] The data based on astronaut photographs from the ISS have been produced by us specifically for this analysis. The procedure is described in detail in A. Sanchez de Miguel 2015, and here we provide a brief review of the key information.

Like all sensors, the Nikon D3s cameras used on the ISS have a limited dynamic range over which the response to light is linear. This range depends on the camera settings (ISO, exposure time, aperture), and on the technology used in the sensor. In the case of the D3s, the linear range is between 100 to 10,000 "adimensional units" (ADU) in the raw file output. The relation between ADU and broadband spectral radiances depends on the settings used in the acquisition. One method to increase the dynamic range of acquisition is to combine different images of the same region taken with different settings. This technique is popularly known as High Dynamic Range Imaging (HDR). In our case, we made a selection of the known images of Madrid,[20] and selected those that provided the most information (Figure 12). These images are photographs of Madrid on the same ISS overpass with different settings. The histograms show the distribution of ADU in the raw images, and it can be seen that in image ISS040-E-081258 the city is overexposed but the outlying regions are in the linear region, while in image ISS040-E-081262 the outlying regions are underexposed but most of the city lights are in the linear region. This pair of images also represent the extreme values of the time of exposure of the set, where all the rest of the settings are the same for the rest of the images.

This technique was used on two pairs of ISS images: ISS030-E-188208 and ISS030-E-188210 from March 28, 2012, and ISS040-E-081262 and ISS040-E-081258 from July 26, 2014 (the sky brightness data are compared to each HDR image separately). The low exposure used for the set ISS040-E-081262 was only possible because the acquisition was made with the help of the Nightpod.[44] This device compensates for the movement of the ISS across the sky, allowing longer exposure times without blurring the images.

The UCM sky brightness survey was used for the calibration of the "World Atlas of artificial sky brightness";[31] figures 17 and 18 and table 3 of that paper demonstrate that the sky brightness of Madrid is similar to that observed in other areas of the world for which observations are available. The impact of the different spectral responses of the different sensors (SNPP/VIIRS/DNB, DMSP/OLS, ISS and SQM) has been discussed by Sanchez de Miguel et al.[22, 25] Although this could produce differences in the observations between the instruments, for stable emission spectra from lights on the ground this could not result in differences larger than 0.5 mag for non-thermal emissions. As a typical streetlight in the region of interest did not change significantly during the 5 years during which the sampling campaigns were undertaken, this effect cannot explain the tail seen in Figure 5. We attribute it instead to the presence of direct lights. Although the spectra of sky brightness depend to some extent on wavelength because of Rayleigh and Mie scattering, the sensors all make broadband observations, and the main light source in the images is sodium lamps, for which the emission spectra is dominated by a strong emission line near 590 nm.

## Acknowledgements

This work was supported by the EMISSI@N project (NERC grant NE/P01156X/1), COST (European Cooperation in Science and Technology) Action ES1204 LoNNe (Loss of the Night Network), the ORISON project (H2020-INFRASUPP-2015-2), the Cities at Night project, the European Union's Horizon 2020 research and innovation programme under grant agreement no 689443 via project GEOEssential, FPU grant from the Ministerio de Ciencia y Tecnologia and F. Sánchez de Miguel.

We acknowledge the support of the Spanish Network for Light Pollution Studies (MINECO AYA2011-15808-E) and also from STARS4ALL, a project funded by the European Union H2020-ICT-2015-688135. This work has been partially funded by the Spanish MICINN, (AYA2016–75808–R), and by the Madrid Regional Government through the TEC2SPACE-CM Project (P2018/NMT-4291). The ISS images are courtesy of the Earth Science and Remote Sensing Unit, NASA Johnson Space Center.

CCMK acknowledges the funding received through the European Union's Horizon 2020 research and innovation programme ERA-PLANET, grant agreement no. 689443, via the GEOEssential project, and funding from the Helmholtz Association

Initiative and Networking Fund under grant ERC-RA-0031. We thank J. Coesfeld for producing Figure 1. We thank the organizers of the LPTMM 2013 conference for providing a stimulating forum in which we discussed the nature of the diffuse light around cities in detail.

## Author contributions statement

A.S.M. and J.Z. conceived the study, A.S.M., J.Z. and J.G. conducted the survey plus other volunteers, A.S.M., J.Z. and C.C.M.K. analyzed the results, A.S.M., C.C.M.K and K.J.G. wrote the original manuscript, A.S.M, J.Z., J.G., C.C.M.K. and K.J.G. conducted the funding requests. All authors reviewed the manuscript.

## Additional information

A.S.M., J.Z., C.C.M.K. and K.J.G. are members of environmental organizations, including Bird Life, Celfosc and the International Dark-Sky Association. A.S.M. occasionally provides consultancy advice for the Instituto de Astrofisica de Andalucia - CSIC and the company SaveStars Consulting S.L.

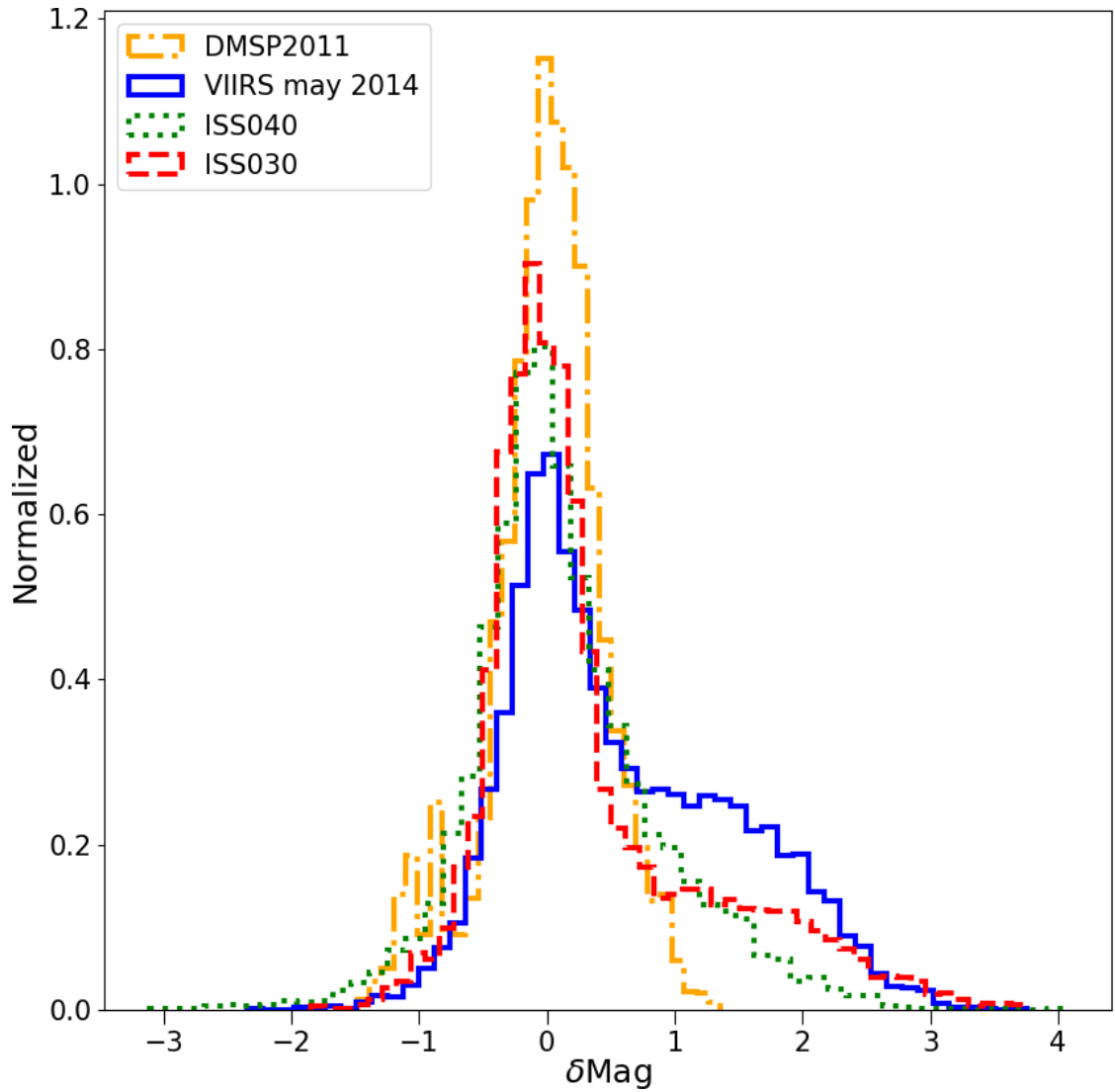

**Figure 5.** Histogram of the difference in observed skyglow radiance from the ground compared to the prediction from the space based observation. Positive values indicate a darker sky than predicted. The imagery is listed in order of decreasing resolution. The histograms have been normalized (the area under the histogram will sum to 1).

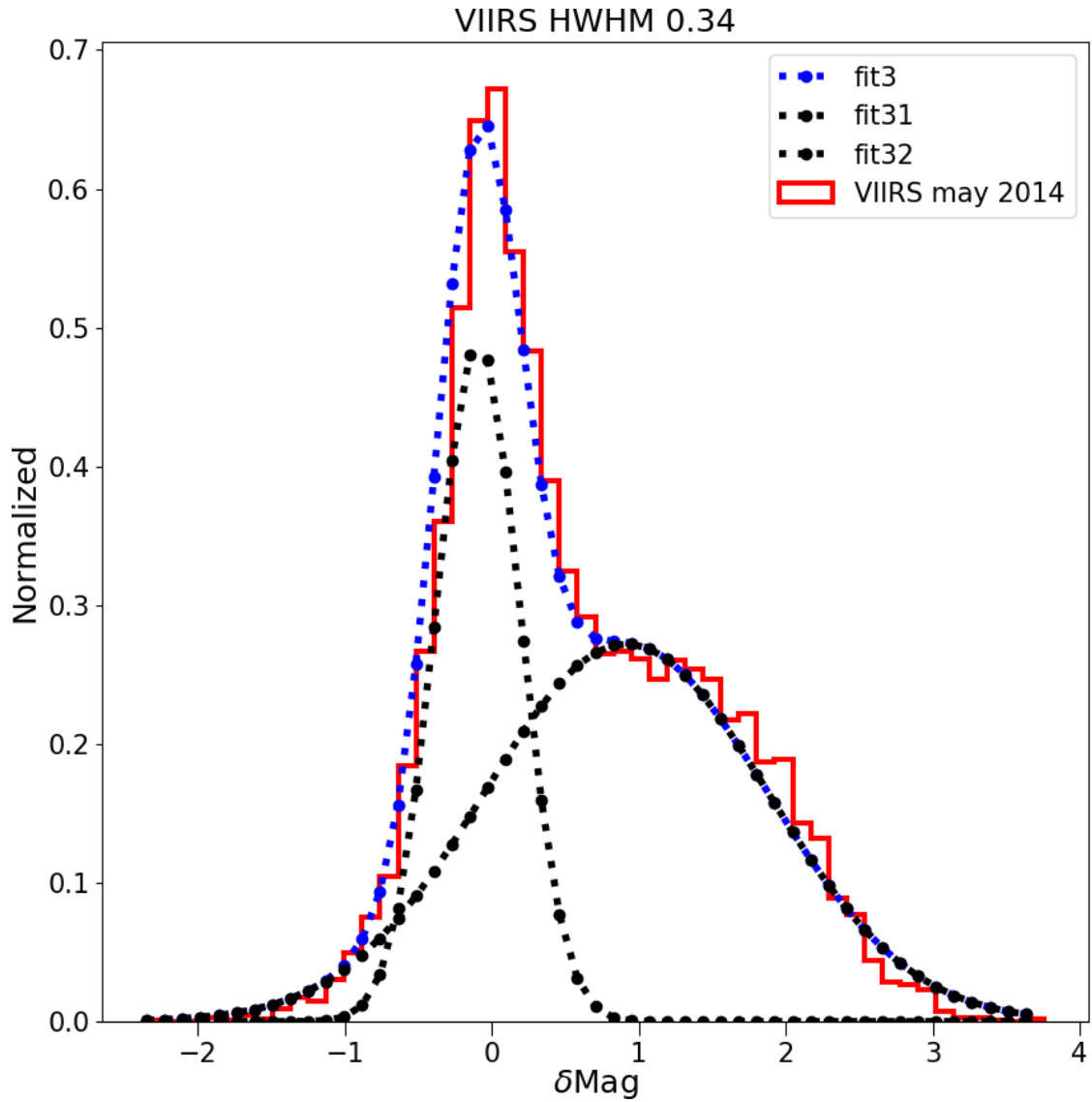

**Figure 6.** Example of double Gaussian fit to the residual of sky brightness for the data from VIIRS DNB. The narrow peak at left represents observations in areas with no installed lighting, the peak at right is areas in which the pixel contains a mix of both sky brightness and light emissions from the ground.

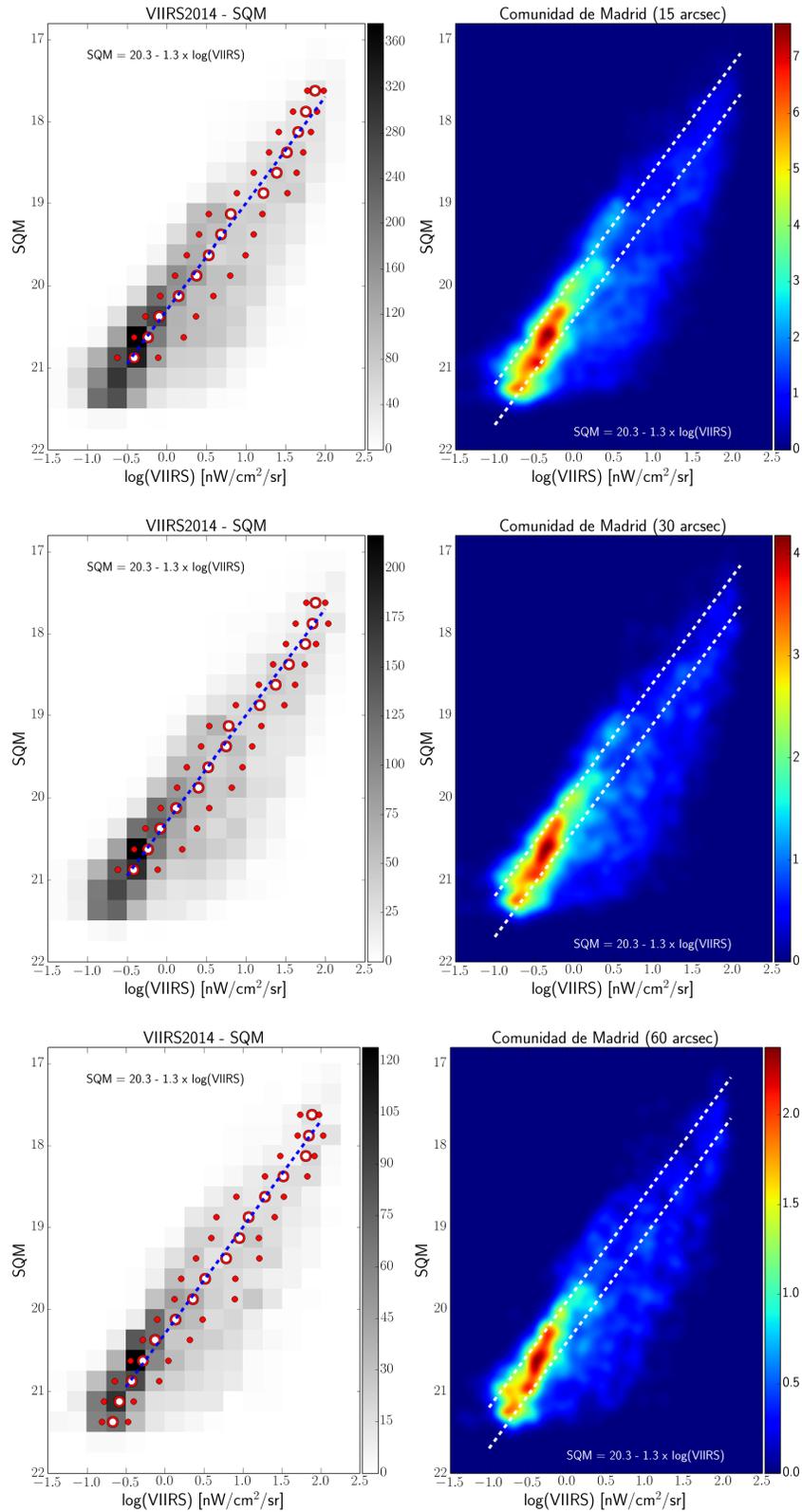

**Figure 7.** Detail about the effect of the sampling size with the sky brightness. 2D sky-brightness distribution with the sampling factor.

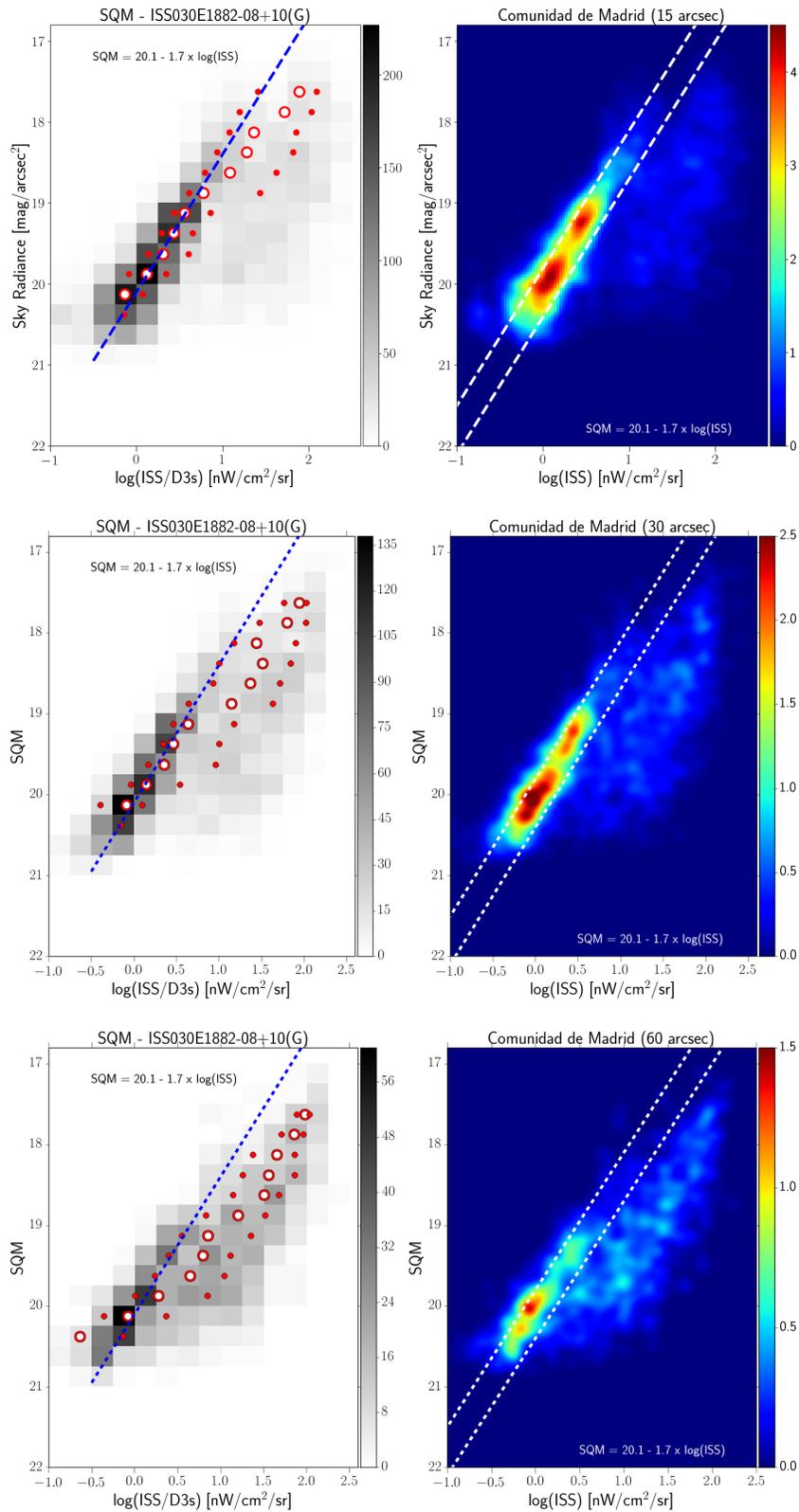

**Figure 8.** Relationship between sky brightness measured at the ground and satellite radiance observation for different sizes of pixel agglomerations. In the top plots, the ISS image data has been averaged over a pixel size of 15" (arcseconds), in the middle plots 30", in the bottom plots 60". In all cases, the relationship is strongest using the high resolution data. In the left plots, the large circle represents the median SQM value, and the small circles represent the 25$^{th}$ and 75$^{th}$ percentile. The right hand plots show the same result using a non-binned color coded density plot with a Gaussian filter. The two dashed lines have the same slope as the blue fit, and are shown for comparison purposes.

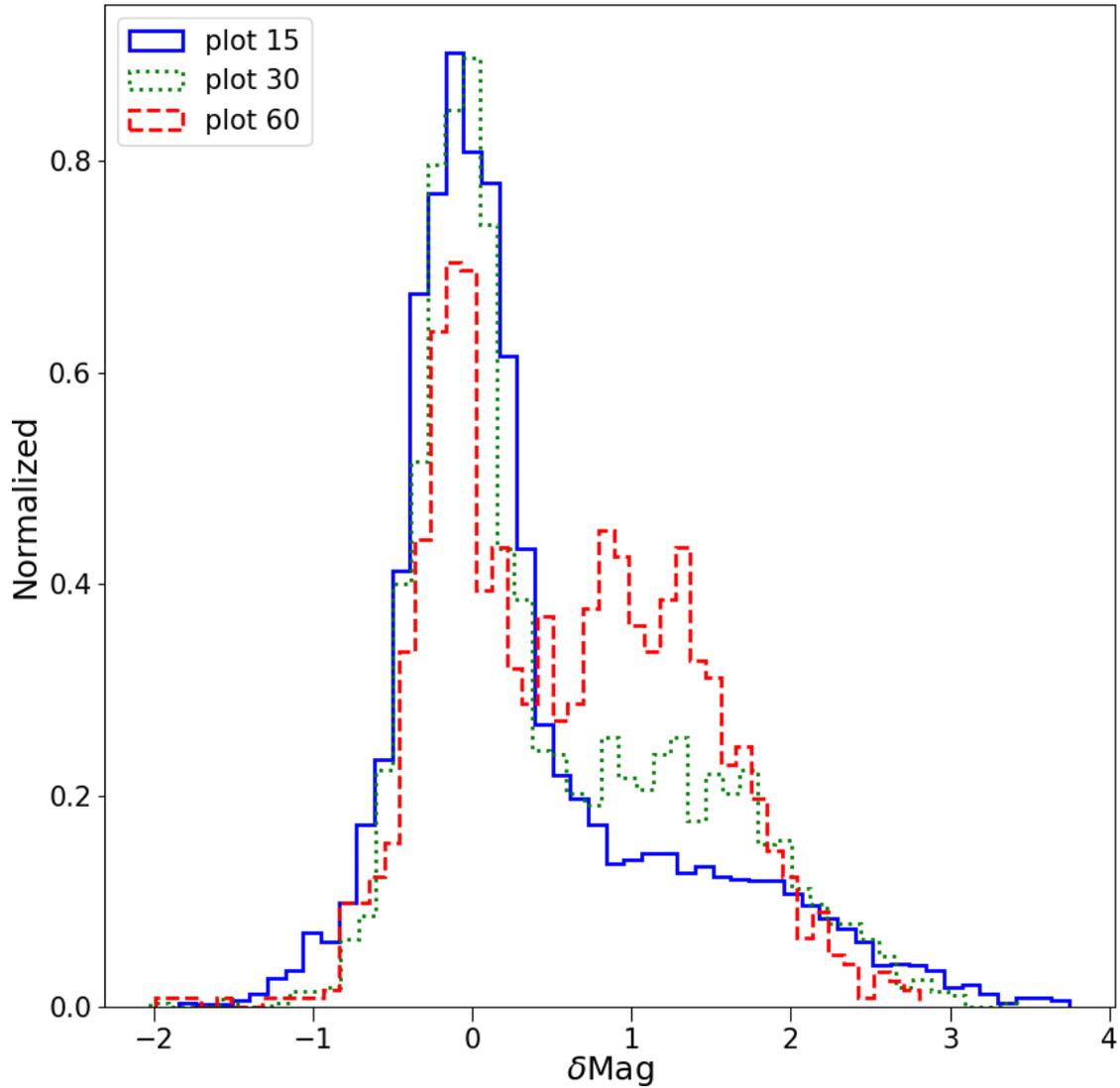

**Figure 9.** Impact of pixel agglomeration area on relationship between sky brightness and radiance detected from space. The histograms show the difference between the observed sky brightness and that predicted based on the ISS photograph. Larger values mean the sky is darker than would be expected. The relationship is best when the pixel size is smallest, or in other words, when the agglomerated pixel is small enough that it is unlikely to contain an admixture of both scattered skyglow and direct emission.

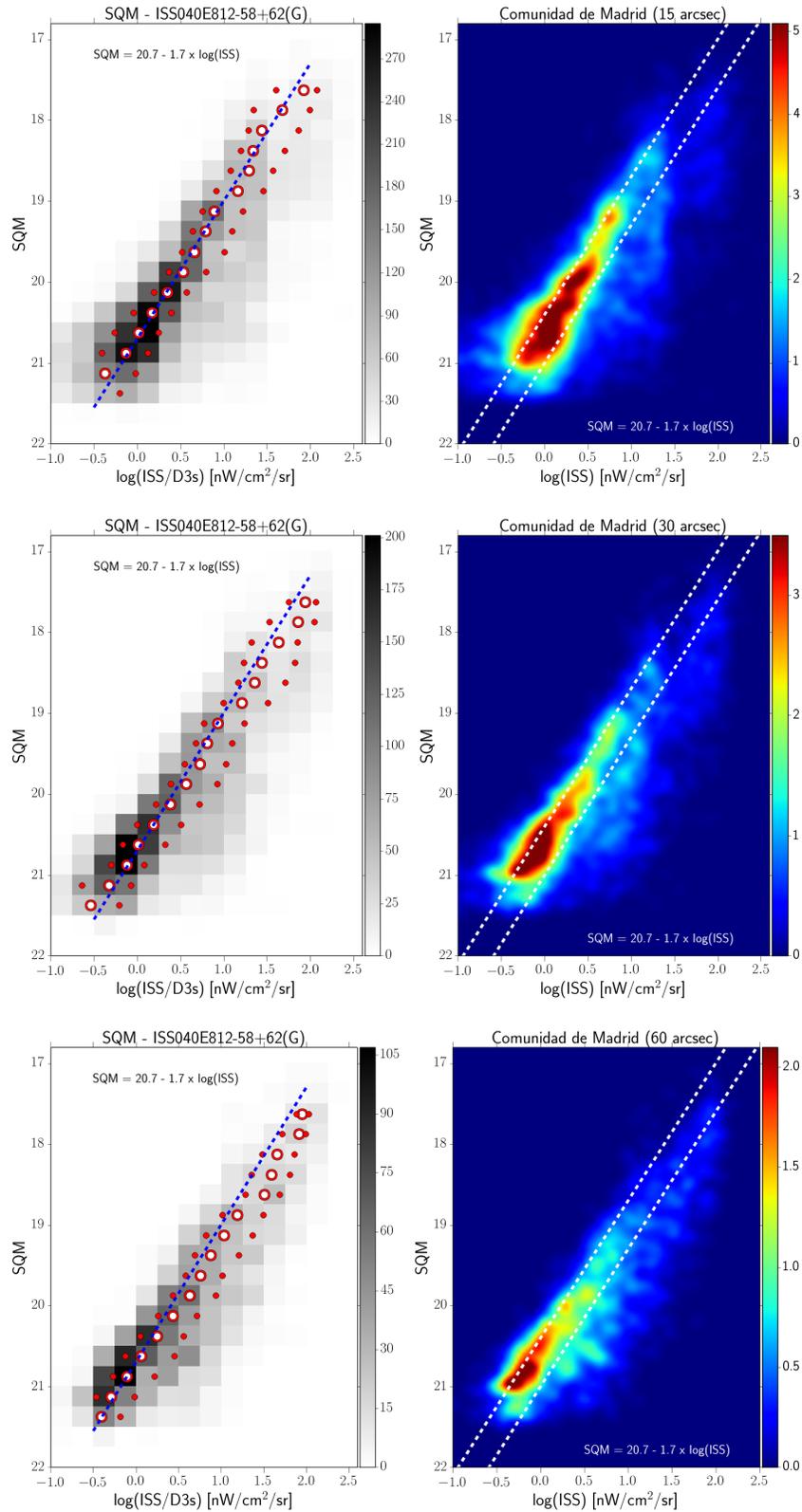

**Figure 10.** Detail about the effect of the sampling size with the sky brightness. 2D sky-brightness distribution with the sampling factor.

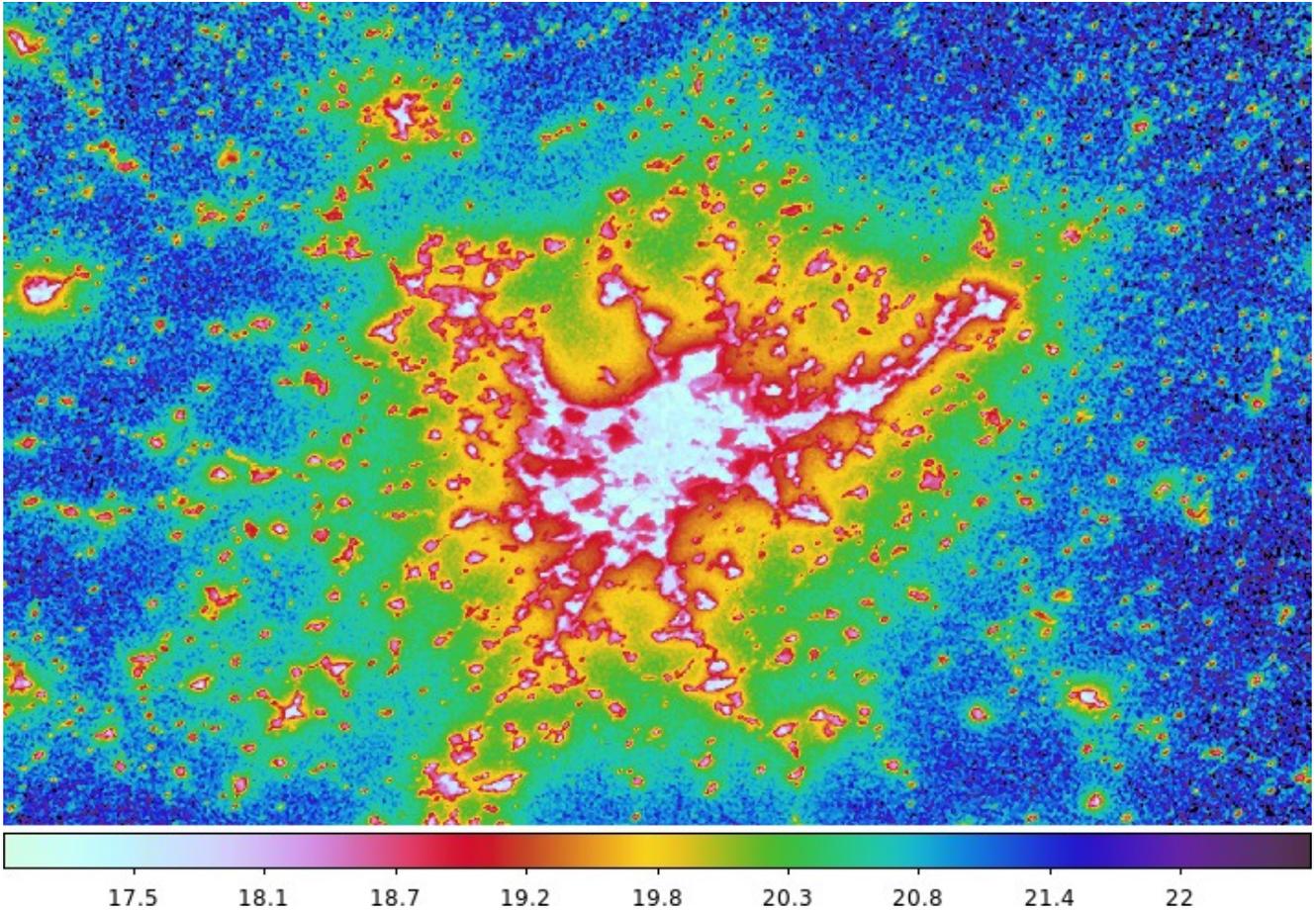

**Figure 11.** Sky brightness map of the central area of Spain, centered on the city of Madrid based on the correlation between the light scattered up and the light scattered down. Scale in *mag/arcsec*$^2$

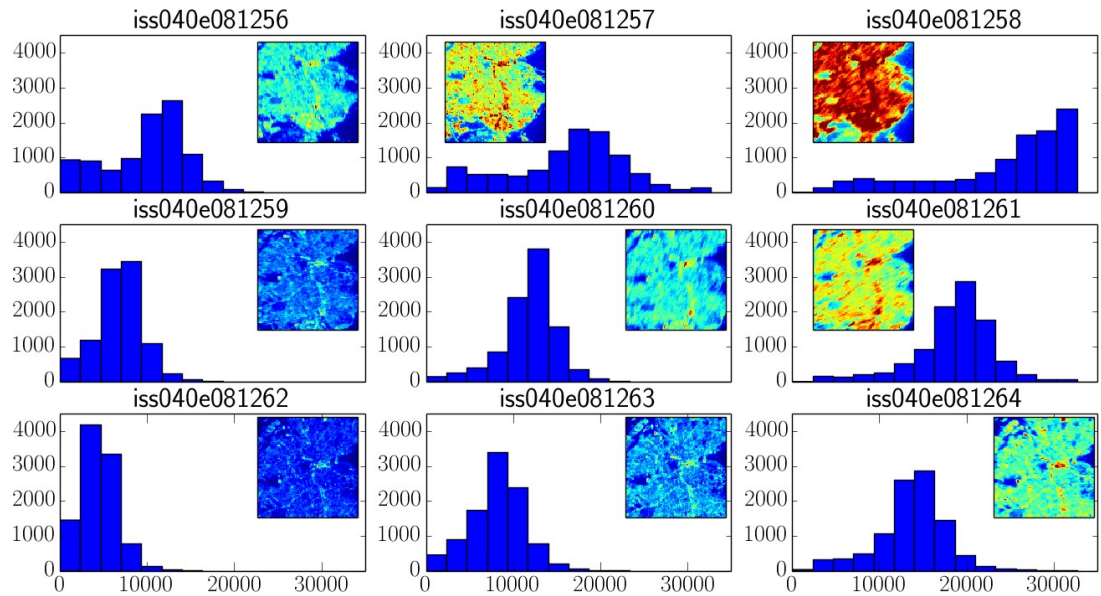

**Figure 12.** A series of histograms for images of Madrid taken using different camera settings by astronauts aboard the International Space Station during a single overpass. The images themselves are shown inset.

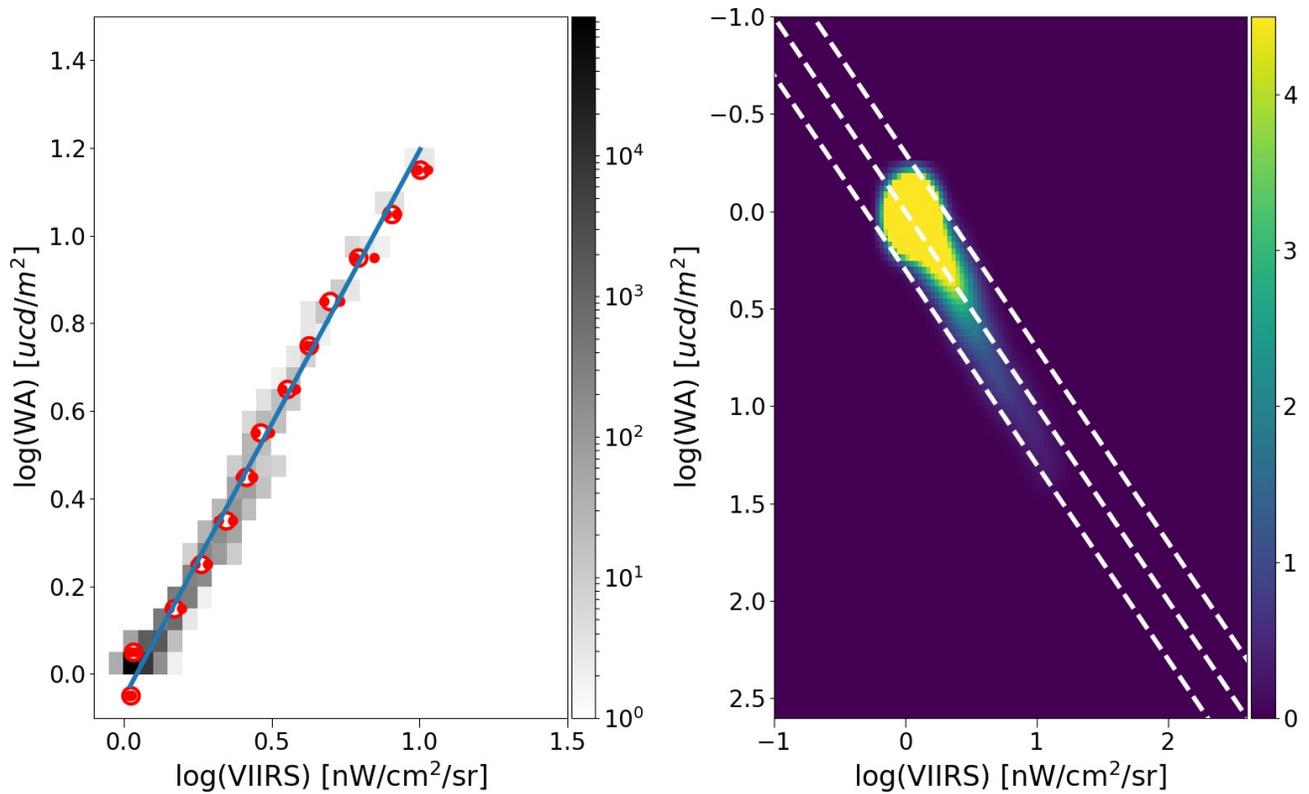

**Figure 13.** Correlation between the World Sky Atlas model[31] and the observed median values from VIIRS. On the left the density plot and median values plus best fit. On the right, line 1:1 and parallel lines with ±0.3.

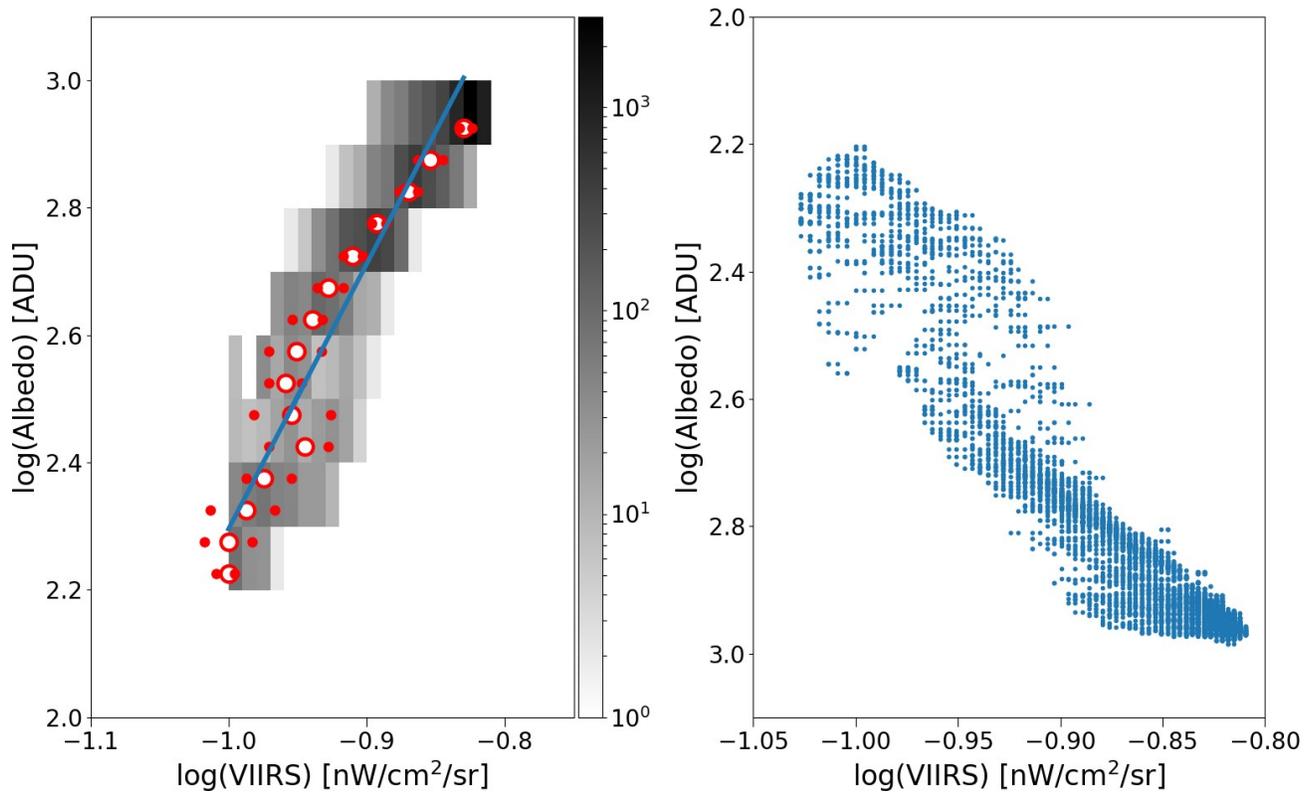

**Figure 14.** Correlation between the MODIS Albedo BSA Bands 1,2,3 and 4 and VIIRS median 2012-2018 in the proximity of the Wal al Namus volcano (Lybia).

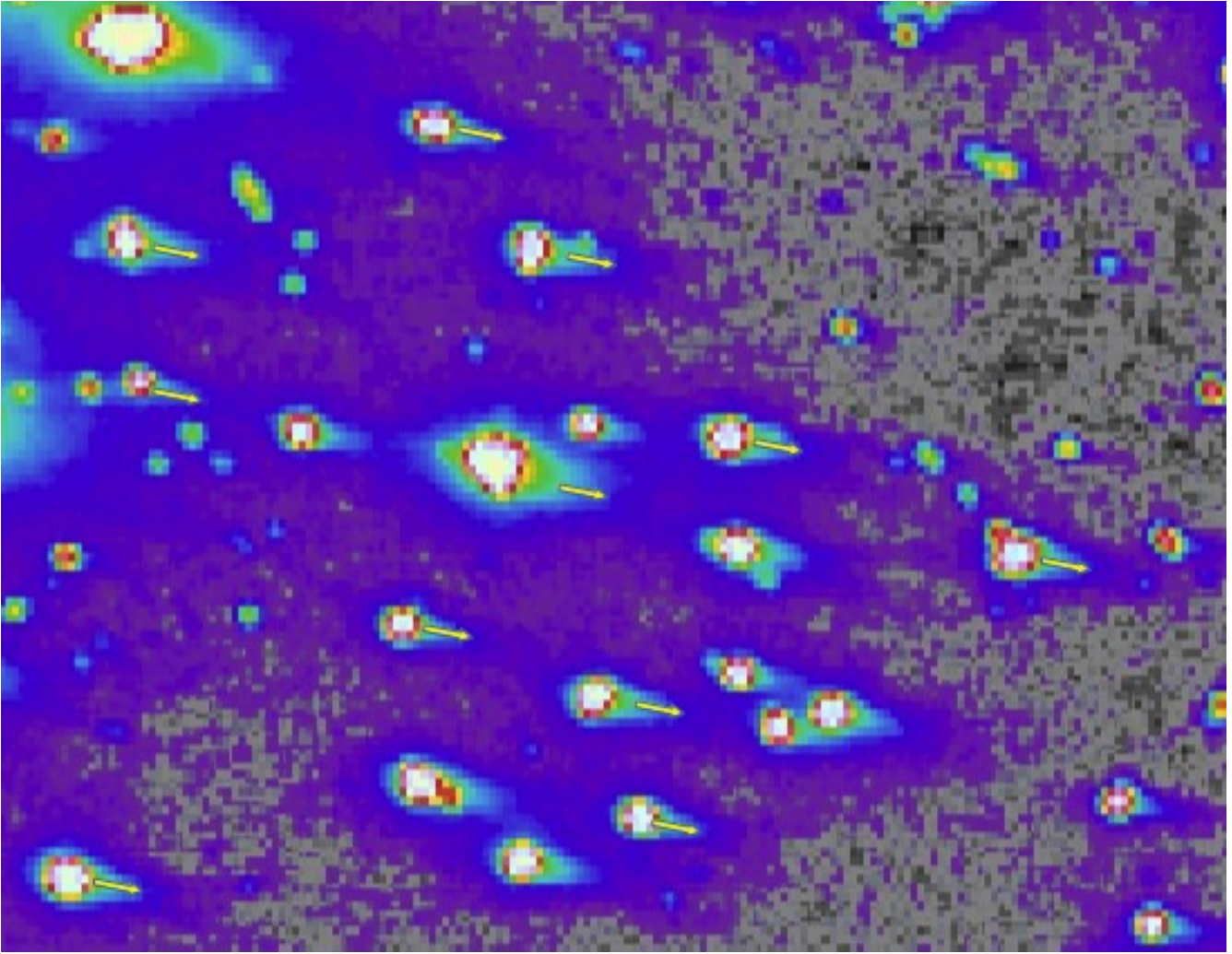

**Figure 15.** Blooming effect at dim sources.